# The massive binary companion star to the progenitor of supernova 1993J

Justyn R. Maund†, Stephen J. Smartt†, Rolf P. Kudritzki‡, Philipp Podsiadlowski*, and Gerard F. Gilmore†

† *Institute of Astronomy, University of Cambridge, Madingley Road, Cambridge CB3 0HA UK*

‡ *Institute for Astronomy, University of Hawaii at Manoa, 2860 Woodlawn Drive, Honolulu, Hawaii 96822 USA*

* *University of Oxford, Astrophysics, Keble Road, Oxford OX1 3RH UK*

**The massive star which underwent core-collapse to produce SN1993J was identified as a non-variable red supergiant star in images of the galaxy M81 taken before explosion[1,2]. However the stellar source showed an excess in UV and B-band colours that suggested it had either a hot, massive companion star or was embedded in an unresolved young stellar association[1]. The spectra of SN1993J underwent a remarkable transformation between a hydrogen-rich Type II supernova and a helium-rich (hydrogen-deficient) Type Ib[3,4]. The spectral and photometric peculiarities were explained by models in which the 13-20 solar mass supergiant had lost almost its entire hydrogen envelope to a close binary companion[5-7]. The binary scenario is currently the best fitting model for the production of such "type IIb" supernovae, however the hypothetical massive companion stars have so far eluded discovery. Here we report the results of new photometric and spectroscopic observations of SN1993J, 10 years after explosion.**

**At the position of the fading SN we detect the unambiguous signature of a massive star, the binary companion to the progenitor. This is evidence that this type of SN originate in interacting binary systems.**

Images of SN1993J were acquired on 28 May 2002 using the Advanced Camera for Surveys (ACS) on board the Hubble Space Telescope (HST). These observations were primarily carried out in the UV and blue spectral regions, where the flux of a hot star companion would be most prominent. A F330W (U-band) image of SN1993J is shown in Figure 1. The high angular resolution of these images (0.025" per pixel) allowed for the complete separation of unrelated surrounding stars from the SN, including the discovery of a previously blended star near to the SN (star G of Figure 1; see Table 1 for details). The blending of stars E-G with the SN (as in previous lower resolution HST images[8]) would lead to the SN appearing bluer. The stars surrounding the SN, which are isolated by the exquisite resolution of the ACS images, would be blended with the SN flux in lower spatial resolution ground-based observations. Recently the contributions of the surrounding blue stars to the U and B band magnitudes of the SN progenitor have been estimated from lower resolution HST archive images[8]. This study concluded that although a significant fraction of the progenitor U-band flux could be explained by the contamination of these stars in the ground-based PSF, within 1σ errors, a hot massive companion star would be compatible with the progenitor and the late-time images. We have conducted a similar analysis with the higher spatial resolution and deeper ACS images. A Gaussian weighting scheme, which is a function of angular distance of the stars from the SN and the wavelength dependent seeing, was applied to simulate the contribution of each object to the photometry of the SN progenitor[1]. We found that these stars, assuming a K0Ia progenitor, would not clearly have had sufficient flux to be solely responsible for the UV excess reported for the progenitor[1,8]. In a similar fashion this photometry provided an estimate of the flux contribution of these surrounding stars to the ground-based spectroscopy of SN1993J.

Spectroscopy of SN1993J was acquired using the LRIS-B spectrometer on the Keck I telescope, covering 3400-5400Å, on 1 March 2003 (Figure 2). The spectrum in general is typical of supernovae at late times which show evidence for the interaction of the ejecta with a dense circumstellar medium[9]. It is dominated by a series of broad, box-like emission lines which have been interpreted to be due to ejecta shocks in a thin spherical shell[10]. However the high signal to noise of the LRIS-B spectrum in the near-UV illustrates two peculiar features not detected before: a rising UV continuum and narrow absorption lines consistent with a hot star spectrum superimposed on the flux from the young SN remnant. The maximum contribution of the surrounding stars to the observed spectrum, again simulating the effects of ground-based seeing conditions with a Gaussian weighting function, was calculated to be 18%. The variability of star A[2,8] (see figure 1), given its magnitude and distance from the SN, has a negligible effect on this analysis. The strength and width of the absorption lines, and the rising shape of the blue continuum suggest that the observed flux is composed of a pseudo-continuum from the SN ejecta emission and flux from a hot star which is unresolved from the SN on the ACS images. The point-spread-function of SN1993J is perfectly consistent with a single stellar source, and conservatively we have estimated that the separation of the star and the supernova must be closer than 0.015 arcseconds (0.3 parsecs), or else we would clearly detect asymmetries in the spatial profile. The observed B-band stellar density of M81 in the vicinity of SN1993J implied a probability of 1 in 2000 of a single randomly located and unrelated star fortuitously lying so close to the SN as to be unresolved.

The narrow stellar lines observed in the SN1993J spectrum were compared with lines from observed spectra of late-O and early B standard stars and TLUSTY[11,12] model stellar spectra, with appropriate parameters[13,14]. The comparison was conducted in the wavelength region of the U-band, least affected by the strong SN emission lines. Main sequence stars were immediately excluded due to their luminosities being too low to produce lines of sufficient strength to match those in the SN1993J spectrum. The strengths of the lines in the SN1993J spectrum, with respect to the continuum, were weaker than those in the stellar spectra due to the continuum of the SN. The continuum of the SN, in units of the B supergiant flux at each HI line, was calculated by adding excess continuum to the stellar spectra until the re-normalised HI line strengths matched those observed in the SN1993J spectrum. In addition the continua ratio was calculated for HeI lines, in order to identify those types which gave consistent matches with the values calculated for the HI lines assuming a normal He abundance. B-type supergiants of type B5 and later were eliminated by this method. Early B-supergiants (B0-4) provided the best consistent fit to the HI and HeI lines. The average ratio of the continua, for all the HI lines, was calculated for each spectral type: B0Ia 1, B1Ia 1.9, B2Ia 1.9, B3Ia 2.3 and B4Ia 3 (where the contribution due to the surrounding stars is already subtracted). From the ACS U and B-band fluxes of the SN plus the companion, and the relative fluxes of the companion required to produce these narrow lines, the luminosity and temperature of the companion star were calculated to be: $\log L/L_\odot = 5 \pm 0.3$ and $\log T_{eff} = 4.3 \pm 0.1$ (our best estimate is a B2Ia star). The uncertainties were determined from the range of atmospheric parameters that would provide appropriate simultaneous matches to the observed lines. The pre-explosion photometry was utilised to further constrain the companion parameters. The luminosity of the B0Ia star requires a U-band magnitude too bright to be compatible with the pre-explosion U-band luminosity of the progenitor binary system. The nature of the progenitor, considering the blend of the progenitor and the companion in the pre-explosion photometry, is restricted to be a K-supergiant with parameters $\log L/L_\odot = 5.1 \pm 0.3$ and $\log T_{eff} = 3.63 \pm 0.05$.

The most important implication of these results is that it confirms that the progenitor of SN1993J was a member of a binary system which has lost most of its hydrogen-rich envelope by its interaction with its companion star[5-7]. Figure 3 illustrates the evolutionary tracks of both components, the mass-losing and the mass-accreting one, in the Hertzsprung-Russell (H-R) diagram. The model is very similar to the ones proposed originally[5-7]. The system initially consists of two stars of comparable masses (15 and 14$M_\odot$) with an orbital period of 5.8 yr. At the time of the supernova explosion the locations of both stars in the H-R diagram are close to the observationally inferred ones. Note that the subsequent evolution of the companion star would be very different from a star evolved in isolation, and it would most likely end its evolution as a blue supergiant rather than a red supergiant, producing a supernova like SN 1987A[15-18] (see figure 3). An overabundance of He would be a signature of the accretion onto the companion from the progenitor and the SN, and we would expect that the He abundance could be determined at much later times once the SN has faded sufficiently[19,20].

The narrow lines in the SN1993J spectrum are at a velocity of $-120$ kms$^{-1}$ relative to the rest frame. This velocity is consistent with the rotational velocity curve of M81 at the position of SN1993J[10,21] and with a low kick velocity to the companion in a long period binary. At the time of the explosion the orbital period of the system is ~25yr, and the companion has an orbital velocity of ~6kms$^{-1}$. The companion is expected to move with this velocity relative to its neighbourhood if the system is disrupted as a result of the supernova, or with the post-supernova binary system velocity (~20 - 40kms$^{-1}$) if the system remains bound[22].

This letter is based on observations made with the NASA/ESA Hubble Space Telescope and the W.M. Keck Observatory, which is operated as a scientific partnership among the California Institute of Technology, the University of California and the National Aeronautics and Space Administration. We thank Robert Ryans for providing us with new TLUSTY calculations of the hydrogen line spectra for B-type supergiants. JRM and SJS thank PPARC for financial support and thank A.D. Mackey and J. Mack for advice concerning photometric data analysis.

Correspondence and requests for materials should be addressed to JRM (e-mail: jrm@ast.cam.ac.uk).

**TABLE 1: Photometric magnitudes of SN1993J and nearby stars**

| Star† | Distance from SN/parsec‡ | Near-UV (F250W)* | | U* | | B* | | V* | |
|---|---|---|---|---|---|---|---|---|---|
| A | 12.9 | 22.99 | (0.05) | 23.05 | (0.08) | 23.78 | (0.02) | 22.56 | (0.01) |
| B | 25.0 | 22.76 | (0.05) | 22.89 | (0.06) | 23.12 | (0.01) | 23.05 | (0.01) |
| C | 21.1 | 22.52 | (0.05) | 22.98 | (0.08) | 23.78 | (0.01) | 23.80 | (0.01) |
| D | 20.8 | 22.15 | (0.06) | 22.87 | (0.07) | 23.84 | (0.01) | 23.95 | (0.01) |
| E | 6.0 | 23.39 | (0.07) | 23.68 | (0.08) | 24.12 | (0.01) | 24.17 | (0.01) |
| F | 5.1 | 23.73 | (0.12) | 24.2 | (0.18) | 25.04 | (0.01) | 24.85 | (0.01) |
| G | 2.8 | 22.79 | (0.19) | 23.54 | (0.08) | 24.44 | (0.02) | 24.42 | (0.01) |
| 1993J | - | 19.93 | (0.04) | 20.7 | (0.05) | 21.08 | (0.01) | 20.27 | (0.01) |

The colours of the surrounding stars imply they are of early-type, in the range O9-B3. Bracketed numbers indicate the uncertainty, in magnitudes, of the preceding magnitude. Photometry of 24 other stars, within 106pc of SN1993J, were utilised to determine the reddening by comparing observed colours (U-B) and (B-V) with intrinsic colours, yielding E(B-V)=0.2 and hence $A_v$=0.62 assuming a standard galactic reddening law, which is in good agreement with previously determined values[10]. An accurate distance to the galaxy of 3.63 ± 0.14 Mpc from Cepheid variables is available[23].

†As identified on Fig 1, utilising the nomenclature of ref. 8.

‡Distances of the stars, in parcsec, calculated from the centre of the SN1993J PSF.

*F250W Vegamag –in the ACS photometric system[24]; U,B and V Johnson magnitudes, transformed from the original ACS photometric system.

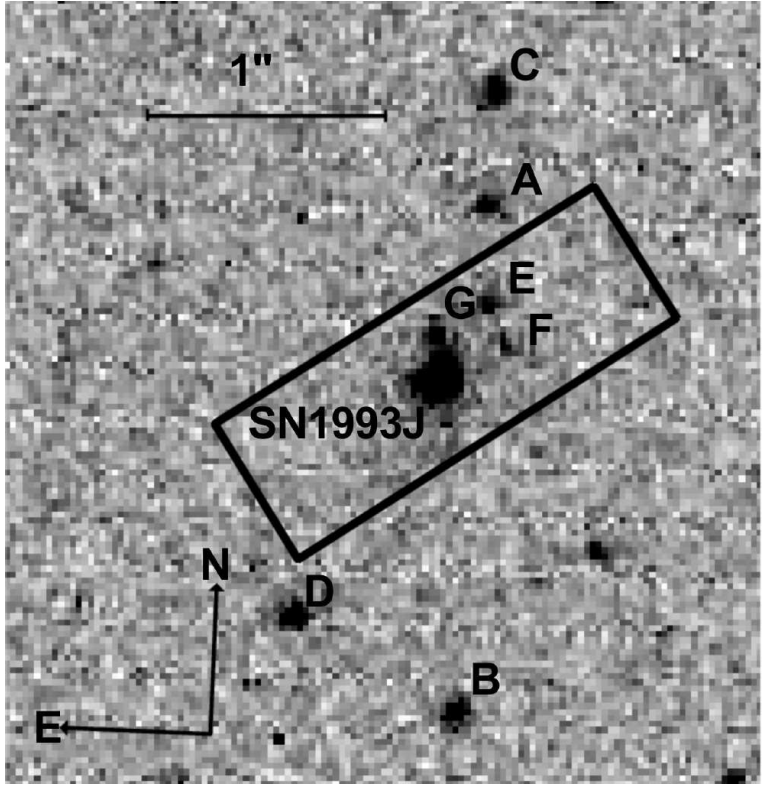

FIG 1: An HST ACS High Resolution Camera (HRC) F330W (~U band) image of SN1993J. The image was a 1200s exposure, acquired on 28 May 2002 - 9.17 years post-explosion, observed for program GO9353. The spatial resolution of the HRC allows the clear separation of stars E-G from the SN (adopting the nomenclature of ref. 8). The photometric magnitudes of the stars measured in the ACS filters[24] are listed in Table 1. These magnitudes were used to calculate, and remove, the contamination of the faint surrounding blue stars to the ground-based spectrum of SN1993J. The box is an example of the area of sky sampled in the lower spatial resolution ground-based spectroscopy. The width is 0.7" (the slit width employed with the LRIS spectrograph), and the length is 1.8" (the aperture diameter used to extract the object spectrum down to a level of 10% of the peak flux). The slit was positioned at the parallactic

angle and during the night rotated between angles -162° and 106° (east of north).  The optical spectrum presented in Figure 2 is effectively a sum of the flux of all objects within this box.

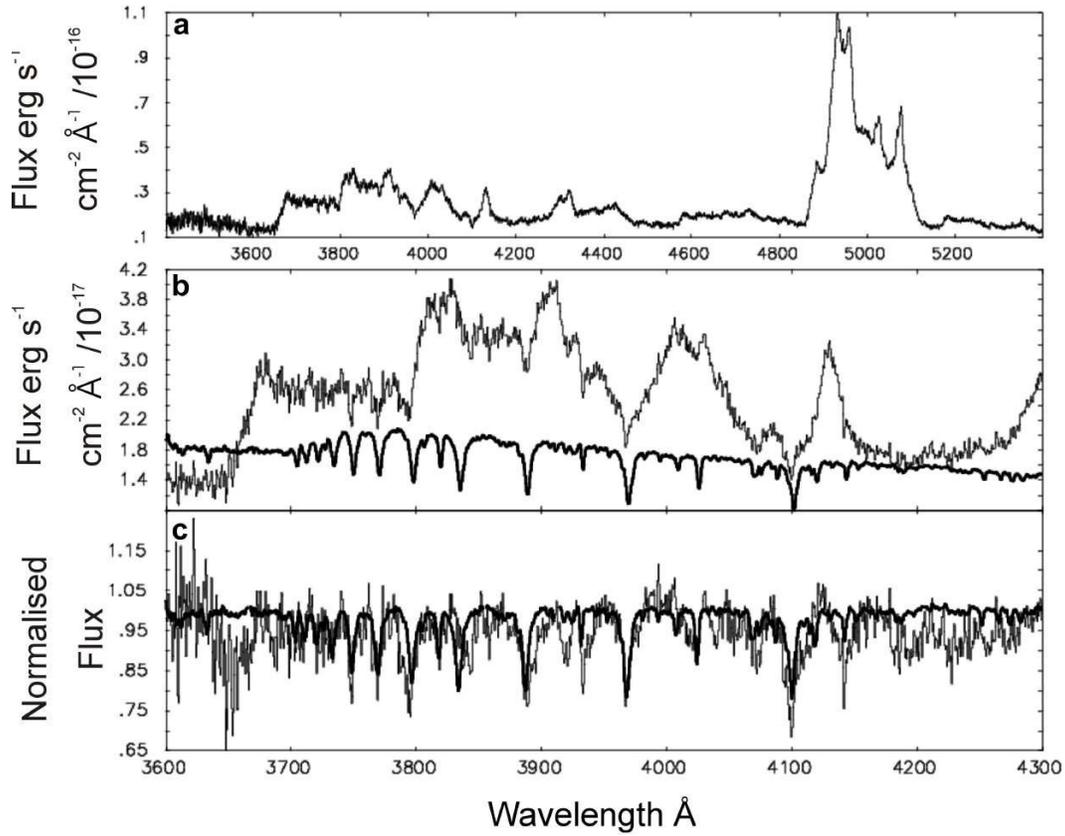

FIG 2: Near UV spectra of SN1993J.

a) Flux calibrated ground-based blue spectrum of SN1993J. The spectrum was acquired with Keck I LRIS-B (1 March 2003, 9.93 years post-explosion) using the 600/4000 grism (with instrumental resolution 2.4Å and signal-to-noise ranging from 15-30). Seeing during the 5.5hr total exposure time was between 0.7-1.2 arcseconds. The spectrum is dominated by broad, box like emission line profiles which are characteristic of interaction between the supernova ejecta and the circumstellar medium[9,10]. However the near UV region shows a series of narrow absorption features.

b) Flux spectra of SN1993J (thin line) and a B-Supergiant (bold line) HD168489 (B1Ia, $T_{eff}$~23.3kK and Log $L/L_\odot$=5). The stellar spectrum was scaled to the distance of M81, with a reddening of E(B-V)=0.2 and a radial velocity shift of -120 kms$^{-1}$. The series of sharp absorption features are coincident with the HI Balmer series, and the HeI line at 3819.7Å. The absorption lines are conspicuously narrower than the [OIII] ($\lambda\lambda$4959, 5007Å) and [SII] ($\lambda\lambda$4068, 4072Å) emission lines (3200kms$^{-1}$) - implying that the absorption lines do not arise from the SN ejecta itself. The flux contribution from the contaminating stars A-G (18%) cannot reproduce the strength of the absorption lines observed.

c) Normalised spectra of SN1993J (thin line) and the B1Ia Galactic standard star HD168489, which has been re-normalised for an excess continuum ratio of 0.8 – in units of the combined continuum fluxes of the companion and surrounding stars. The SN continuum was normalised by estimating the position of the continuum and applying a series of short spline fits. The HI Balmer and HeI absorption lines are best matched by early-B type supergiant spectra (B0-B4). The average ratios, in the range 3600-4000Å, and the total measured ACS flux in the F330W filter allowed a calculation[14] of the parameters of the unresolved hot star component: $logL/L_\odot$=5±0.3, $logT_{eff}$=4.3±0.1.

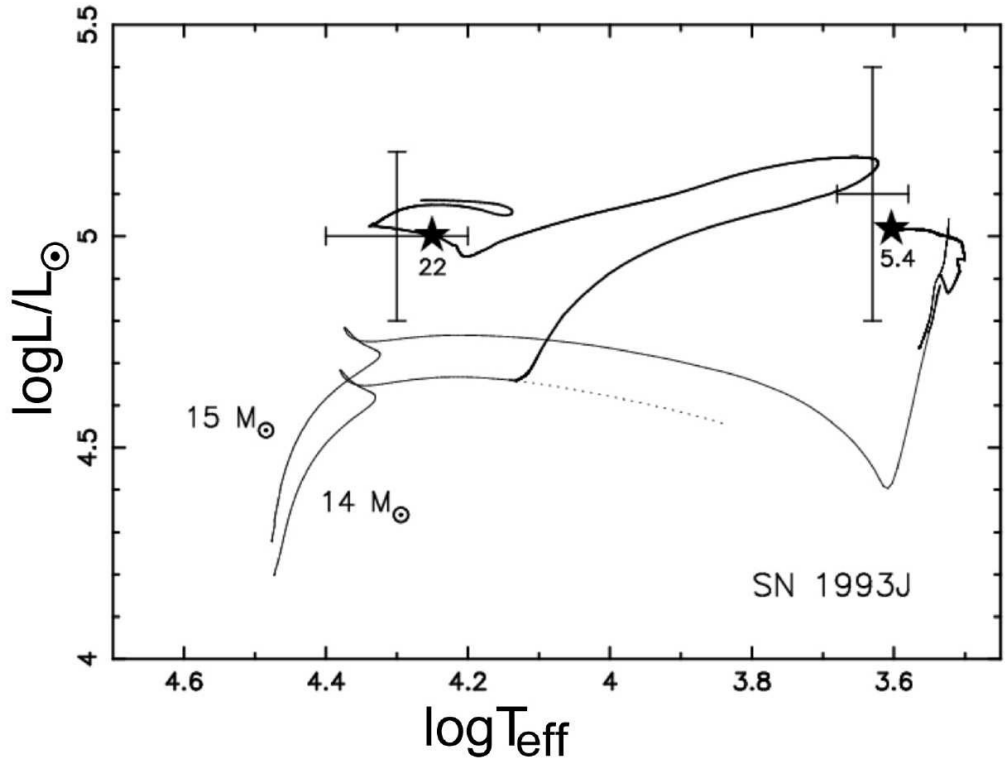

FIG 3: Hertzsprung-Russell diagram illustrating the evolution of the progenitor binary of SN1993J. The system initially consists of two stars of 15$M_\odot$ and 14$M_\odot$ in an orbit of 5.8 yr. The thin lines show the evolution of both components before the beginning of mass transfer, the bold lines during the mass transfer phase. The more massive component fills its Roche lobe and starts to transfer mass to the companion after the exhaustion of central helium when ascending the red-supergiant branch for the second time (so-called case C mass transfer; alternatively mass transfer may already start on the first ascent of the red-supergiant branch; so-called case BC mass transfer). The mass transfer rate is initially very high, reaching a peak of $4\times10^{-2}$ $M_\odot yr^{-1}$ (the accretion rate onto the secondary was limited to $2\times10^{-2}$ $M_\odot yr^{-1}$). Once the primary has transferred most of its envelope, it detaches from its Roche lobe and subsequently continues to lose mass due to a strong stellar wind, assumed to be $4\times10^{-5}$ $M_\odot yr^{-1}$ (ref. 25) at the time of the explosion. Because of the similar masses, the secondary is already somewhat evolved and is near the end of its hydrogen core burning phase or just beyond (and hence is not being rejuvenated by accretion[15,16]). The numbers along the tracks give the masses of the two components at selected phases. At the time of the explosion of the primary (indicated by asterisks), the primary has a mass of $5.4M_\odot$ (with a helium-exhausted core of $5.1M_\odot$), the secondary has a mass of $22_\odot$. The error bars show the parameters of the two components, as determined in this work.